\documentclass[preprint,12pt,number]{elsarticle}

\usepackage{amsmath,amssymb}
\usepackage{graphicx}
\usepackage{booktabs}
\usepackage[T1]{fontenc}
\usepackage[utf8]{inputenc}

\journal{arXiv}

\begin{document}

\begin{frontmatter}

\title{Boundary-condition identification from a scale-invariant
frequency ratio in shear-deformable and nonlocal beams and plates}

\author{Ak\i{}n Oktav}
\address{Department of Mechanical Engineering, Alanya Alaaddin Keykubat
University\\Alanya 07425 Antalya, T\"{u}rkiye \\ E-mail: akin.oktav@alanya.edu.tr, ORCID: 0000-0001-5983-3953}

\begin{abstract}
The ratio of a structure's first two natural frequencies depends on its boundary
conditions but not on its size, material density, or absolute stiffness,
which multiply every frequency equally and cancel in the quotient. We use this to
identify the boundary conditions of a small-scale beam or plate from two ratio
measurements, without knowing its exact dimensions or material. The paper first
establishes how the ratio $r=\omega_2/\omega_1$, and the fractional shift $\Delta r$
it undergoes when a boundary condition changes, respond to the two effects that
matter at small scale: transverse shear deformation, through Timoshenko beam and
Mindlin plate theory, and small-scale elasticity, through Eringen's nonlocal
Timoshenko model. Using a locking-free Rayleigh--Ritz beam solver and a
finite-element Mindlin plate solver, each checked against benchmarks, we find an
asymmetry. The absolute ratio moves appreciably with thickness, by 12 to 14
per cent over the shear-deformable range, and with orthotropy, by 6 to 9 per cent,
but only weakly with nonlocality; the shift $\Delta r$, by contrast, varies by less
than 0.2 per cent over the full nonlocal range and by a few per cent under
thickness change. Because $\Delta r$ is largely independent of these details, a
measured shift identifies which boundary-condition transition a deliberate change
of support has produced, with a resolution we quantify and verify against
molecular-dynamics data. Two spring continua then extend the method to estimating
support compliance, with the non-uniqueness of the tip-support case stated
explicitly. The method needs no value of the small-scale parameter or the material
constants.
\end{abstract}


\end{frontmatter}

% ============================================================
\section{Introduction}
\label{sec:intro}

The natural frequencies of a slender elastic structure are among the most
reliably measured quantities in mechanics, and among the most informative. A
frequency spectrum encodes the stiffness distribution, the inertia distribution,
and the constraints imposed at the boundaries. The inverse use of vibration
data, that is, inferring a structure's properties or its support conditions from
its measured spectrum, underlies modal testing, structural health monitoring
\citep{Madinier2025}, and the characterisation of micro- and nanoscale resonators
\citep{Ewins2000}. In each of these settings the measured frequencies depend on a
long list of physical parameters, and separating the quantity of interest from
the rest is the central difficulty.

A dimensionless frequency ratio offers a partial escape from this difficulty. The
ratio of the second to the first natural frequency, $r=\omega_2/\omega_1$, is
independent of the absolute scale of the system, of the material density, and of
the overall stiffness, since these factors multiply every eigenfrequency equally
and cancel in the quotient. For a one-dimensional beam governed by the classical
Euler--Bernoulli equation, $r$ is a pure number fixed entirely by the boundary
conditions, for example $r\approx 4.00$ for a simply supported beam and
$r\approx 2.76$ for a doubly clamped beam \citep{Blevins1979}. The fractional
change in this ratio when a boundary condition is altered,
$\Delta r=[r(B_2)-r(B_1)]/r(B_1)$, therefore tracks the boundary condition while
being independent of the size, density, and stiffness of any particular specimen. Whether this indifference survives the
physical corrections that become important at small scales is the question this
paper addresses.

Two such corrections matter. The first is transverse shear deformation. The
classical thin-plate (Kirchhoff) and thin-beam (Euler--Bernoulli) theories
neglect shear, which is acceptable for slender, thin structures but fails as the
thickness-to-length ratio grows. First-order shear deformation theory (FSDT)
restores the shear degree of freedom by allowing cross-sections to rotate
independently of the mid-surface slope. It was introduced for plates by
\citet{Reissner1945} for static bending and by \citet{Mindlin1951} for dynamics,
and its one-dimensional counterpart is the Timoshenko beam. Shear deformation
softens the higher modes more than the fundamental, and so it changes the
spectral ratio. The second correction is small-scale, or nonlocal, elasticity. At
nanometre scales the classical assumption that stress at a point depends only on
the strain at that point breaks down, because interatomic forces are inherently
long-ranged. Eringen's nonlocal theory captures this by making the stress at a
point depend on the strain field in a neighbourhood, governed by a small-scale
parameter $e_0a$ \citep{Eringen1972,Eringen1983}. The application of nonlocal
continuum models to nanostructures was established by
\citet{Peddieson2003}, and exact solutions of the nonlocal Euler--Bernoulli and
Timoshenko beams were later given by \citet{Tuna2016}. Closed-form nonlocal
solutions have since been extended to plate vibration and benchmarked
against molecular dynamics \citep{Han2025}.

Boundary conditions are treated throughout as a discrete set of categories, and
it is useful to name them at the outset. For the beam four ideal supports are
considered: simply supported (SS), clamped at both ends (CC), clamped-free or
cantilever (CF), and clamped-pinned (CP). For the square plate the condition on
each of the four edges is simply supported (S) or clamped (C), and three
combinations are studied, SSSS, SCSC and CCCC. Real supports are rarely ideal,
so two continua are also introduced that connect the categories: a rotational
spring at the ends of a pinned beam, which carries it from SS to CC, and a
translational spring at the tip of a cantilever, which carries it from CF to CP.
A third perturbation, material orthotropy, is included alongside shear
deformation and nonlocality because many small-scale plates, whether layered,
fibrous or crystalline, are anisotropic, and a boundary-condition diagnostic is
of little use if it is confounded by a material property that is itself hard to
measure.

The vibration of single-walled carbon nanotube (SWCNT) resonators is the physical
setting in which both corrections are simultaneously relevant, and it has been
studied extensively. Nonlocal beam and shell models have been used to compute how
the natural frequencies of nanotubes depend on the small-scale parameter, the
length-to-diameter ratio, and the boundary conditions, with the consistent
finding that increasing the nonlocal parameter reduces the frequencies and that
the effect is most pronounced for short tubes and higher modes
\citep{Chwal2018,Bocko2014}. The nonlocal Timoshenko approach to nanobeam vibration was developed early by
\citet{Murmu2023}, and the framework remains an active line of work, with recent
studies extending it and the related two-phase and strain-gradient theories to
micro-structured, functionally graded and sandwich beams
\citep{Xu2025,Dang2026}. Molecular-dynamics studies have established the
bending behaviour of clamped and cantilevered nanotubes and benchmarked the
continuum beam models against atomistic results \citep{Zhang2009}. The
experimental platform for resolving multiple flexural modes of suspended
nanotubes is mature, through electrical mixing spectroscopy and related
techniques, with quality factors reaching several million at low temperature
\citep{Sazonova2004,Moser2014}.

Across this large body of work, the reported quantity is almost always either the
absolute frequency or the ratio of a nonlocal frequency to its local counterpart
at a fixed mode number. The behaviour of the cross-mode spectral ratio $r$, and
in particular of the boundary-condition shift $\Delta r$, has not been examined as
an object in its own right, and the question of whether $\Delta r$ is stable
against shear, nonlocal, and material perturbations has not, to the author's
knowledge, been posed. This is the gap the present paper fills, and the four
contributions are stated here directly.

First, using a nonlocal Timoshenko formulation solved on a locking-free
Rayleigh--Ritz basis and validated against an exact closed-form solution, against
the Euler--Bernoulli thin-beam limits, and against the isogeometric benchmark of
\citet{Lee2013}, two behaviours that are usually reported together are separated:
the absolute ratio softens by about 13 per cent with slenderness and carries the
full nonlocal signature, whereas the shift $\Delta r$ varies by two per cent across
the same slenderness range and by less than 0.2 per cent across the full
literature range of the small-scale parameter. This is the first contribution: at
the nanoscale the ratio and the shift carry distinct physical information.

Second, the boundary-condition shift is developed into a method. Which
boundary-condition transition a deliberate change of support has produced is
identified from two ratio measurements, four frequencies in all, made on the same
platform before and after the change, by matching the shift to a computed
reference library; the resolution of this inverse use of the spectrum is
characterised explicitly and shown to exceed, by roughly an order of magnitude,
what the absolute ratio allows when the scale and material are unknown. The
procedure is demonstrated on molecular-dynamics data, where it assigns the
boundary-condition class from the spectrum alone. This is the second contribution:
the shift becomes a practical, parameter-free way to identify a boundary-condition
change, with a stated resolution.

Third, the discrete library is extended to two continua of real supports, a
rotational spring at the ends of a pinned beam and a translational spring at the
tip of a cantilever, so that the shift reports not only which support condition
holds but how compliant it is. The nonlocal invariance survives along both
continua. The two differ in one respect that is stated plainly: the rotational
continuum inverts uniquely, whereas the tip-support continuum passes through a
minimum, so that a band of measured shifts corresponds to two stiffnesses. This is
the third contribution: a continuous estimator of support compliance, with its
domain of uniqueness made explicit.

Fourth, the same programme is carried to the two-dimensional case. Using a finite
element Mindlin plate solver whose spectral ratios reproduce the
\citet{Leissa1969} thin-plate benchmarks to better than 0.1 per cent, the ratio
and the shift are computed for simply supported, mixed and clamped square plates
across the shear-deformable range of thickness and under strong material
orthotropy. The absolute ratio responds to both, by 12 to 14 per cent and by 6 to
9 per cent respectively, while the shift stays within about ten per cent under
each. This is the fourth contribution: the asymmetry between ratio and shift is a
property of the operator, not of the beam, and it holds against a material
perturbation that has no one-dimensional counterpart.

The reason $\Delta r$ is so stable is a double cancellation. The leading shear and
nonlocal corrections enter the absolute
frequencies in a nearly mode-independent multiplicative way, so they cancel in
the ratio and cancel again in the boundary-condition difference, leaving a
fingerprint of the boundary condition alone. This property has a direct practical
value, since it makes $\Delta r$ a robust diagnostic for identifying or verifying
boundary conditions in micro- and nanoscale resonators, one that does not require
prior knowledge of the small-scale parameter, the precise thickness, or the full
set of material constants. A confirmation of the same mechanism appears within our
own validation: the plate spectral ratio reproduces the Kirchhoff benchmark to a
tenth of a per cent even where the absolute frequencies from the same element
differ by about one per cent, so the ratio cancels common-mode discretisation
error exactly as it cancels the physical contributions.

The remainder of the paper is organised as follows. Section~\ref{sec:ratio}
defines the spectral ratio and its shift. Section~\ref{sec:beam} sets out the
nonlocal Timoshenko beam model, its locking-free discretisation, and its
validation. Section~\ref{sec:beamresults} reports the beam results: the dependence
on slenderness, the separation of the nonlocal signature in the absolute
frequencies from the near-invariance of the ratio, the robustness of the shift,
and the comparison with published nanoscale data. Section~\ref{sec:inverse}
develops the shift into an inverse method for boundary-condition identification,
states the measurement protocol, characterises the resolution, and extends the
method along the two support continua. Section~\ref{sec:plates} carries the
programme to the shear-deformable plate, including material orthotropy.
Section~\ref{sec:discuss} discusses why the shift is scale-invariant and the
limitations of the present treatment, and Section~\ref{sec:concl} concludes.

\section{The spectral ratio and the boundary-condition shift}
\label{sec:ratio}

Let a linear elastic structure undergo free harmonic vibration, so that its
natural angular frequencies $\omega_1<\omega_2<\dots$ are the square roots of the
eigenvalues of the governing operator under a prescribed set of boundary
conditions. The quantity of interest is the spectral ratio of the first two
flexible modes,
\begin{equation}
r(B) = \frac{\omega_2(B)}{\omega_1(B)},
\end{equation}
where $B$ denotes the boundary-condition set. Because every eigenfrequency scales
as $\omega\propto\sqrt{\text{stiffness}/\text{inertia}}$, any factor multiplying
the stiffness or the inertia uniformly across modes cancels in the quotient. For a
homogeneous beam or plate this removes the dependence on overall size, on material
density, and on the absolute elastic modulus, so $r$ is fixed by the shape of the
governing operator and by $B$ alone. The central object is the fractional change
in $r$ produced by a change of boundary condition from $B_1$ to $B_2$,
\begin{equation}
\Delta r = \frac{r(B_2)-r(B_1)}{r(B_1)}.
\end{equation}

% ============================================================
\section{The nonlocal Timoshenko beam: model, discretisation and validation}
\label{sec:beam}

\subsection{Nonlocal Timoshenko model for the nanoscale}
\label{sec:nonlocal}
At the nanoscale the local constitutive relation is replaced by Eringen's nonlocal
model, in which the stress at a point depends on the strain field in a
neighbourhood \citep{Eringen1972,Eringen1983}. In differential form the relation
carries a single small-scale parameter $e_0a$, and the one-dimensional flexural
problem is the nonlocal Timoshenko beam \citep{Tuna2016}. Alternative size-dependent
formulations of the Timoshenko beam, such as micromorphic and strain-gradient
models with more than one length scale, continue to be developed
\citep{Challamel2024,Timtaoucine2026,Xu2025};
the present work adopts the differential Eringen form as the most economical vehicle
for the spectral-ratio question. In the variational form
used here, nonlocality augments the mass operator: each inertia term
$\int \phi_i\phi_j$ is replaced by $\int(\phi_i\phi_j+\mu\,\phi_i'\phi_j')$, the
standard nonlocal mass augmentation, which reduces to the local Timoshenko beam as
$\mu\to 0$, with $\mu=(e_0a/L)^2$ and $L$ the beam length. Frequencies are
reported as $\lambda^2=\omega L^2\sqrt{m/EI}$, following the convention of
\citet{Lee2013}, where $m$ is the mass per unit length, so that the slender simply
supported beam gives $\lambda_n=n\pi$ in the local Euler--Bernoulli limit.

The dependence of the spectral ratio on the tube geometry enters only through a
single dimensionless slenderness group. For a Timoshenko beam the relative weight
of the shear and rotary-inertia corrections is set by the radius of gyration of
the cross-section, $r_g=\sqrt{I/A}$, through the ratio $r_g/L$. A single-walled
carbon nanotube is therefore mapped to the beam model by taking
$r_g=d/(2\sqrt{2})$ for a thin-walled cylinder of diameter $d$, a purely geometric
quantity. This avoids any appeal to an effective wall thickness, a quantity for
which reported values for single-walled nanotubes vary by roughly a factor of five
depending on the calibration adopted. Because the spectral ratio depends on the
geometry only through $r_g/L$, and because the dimensional shell parameters cancel
in the ratio, the present results are insensitive to that unresolved question by
construction. The nonlocal parameter $e_0a$ is swept across the range 0 to 2~nm
reported in the literature \citep{Chwal2018}, rather than fixed at a single
calibrated value, so the nanoscale results are presented as a sensitivity across
that range.

\subsection{Discretisation and boundary conditions}
\label{sec:beamdisc}
The nonlocal Timoshenko problem is solved by the Rayleigh--Ritz method. The
deflection $w$ is expanded in the characteristic functions of the corresponding
local Euler--Bernoulli problem, sine functions for the simply supported beam and
the clamped-clamped or clamped-free beam functions otherwise, and the rotation
$\psi$ is expanded in the derivatives of the same functions. This consistent
choice matters. The shear energy penalises $w'-\psi$, so a basis in which $\psi$
cannot represent $w'$ exactly locks in the slender limit and overstiffens the
clamped modes, the same shear locking that Timoshenko finite elements avoid by
hierarchical or consistent interpolation \citep{Dang2026}; with $\psi$ drawn from
the derivatives of the $w$ basis the
constraint $w'=\psi$ is exactly representable and the thin-beam limit is
recovered cleanly. Ten terms are used per field, and the hyperbolic beam functions
are evaluated in a form that avoids the cancellation of large terms at high order.
For the simply supported beam the discrete problem reduces, mode by mode, to a
$2\times 2$ generalised eigenvalue problem solvable in closed form, providing an
exact reference.

Elastic supports are represented by springs added to the stiffness matrix, the
standard device for arbitrary boundary conditions in Ritz treatments of Timoshenko
beams \citep{Jin2024}. A
rotational spring of stiffness $\kappa_\theta$, in units of $EI/L$, at each pinned
end contributes $\tfrac12\kappa_\theta[\psi(0)^2+\psi(1)^2]$ and carries the beam
from simply supported to clamped; a translational spring of stiffness $\kappa_w$,
in units of $EI/L^3$, at the free end of a cantilever contributes
$\tfrac12\kappa_w\,w(1)^2$ and carries it from clamped-free to clamped-pinned.
Both are used in Section~\ref{sec:inverse}. The beam boundary conditions
considered are therefore simply supported (SS), clamped-clamped (CC), clamped-free
(CF), and clamped-pinned (CP), the last reached as the stiff limit of the tip
spring.

\subsection{Validation}
\label{sec:beamvalid}
The beam solver passes three checks before any result is reported. First, for the
simply supported beam it reproduces the exact mode-by-mode $2\times 2$ solution to
better than one part in $10^{10}$ across the full range of thickness and nonlocal
parameter. Second, in the slender local limit ($h/L=0.005$, $e_0a=0$) the
clamped-clamped and clamped-free spectral ratios approach the Euler--Bernoulli
values $2.7565$ and $6.2669$ from below, to within $0.02$ per cent, and the
clamped-pinned ratio $3.2406$ is recovered to $0.004$ per cent as the stiff limit
of the tip spring. Approach from below is the signature of a locking-free basis;
a locked basis overshoots the thin-beam limit. Third, for the clamped-clamped
Timoshenko beam at $h/L=0.01$ the present frequency parameters
$\lambda_1=4.7285$ and $\lambda_2=7.8479$ agree with the converged isogeometric
values $4.7284$ and $7.8469$ of \citet{Lee2013} to $0.002$ and $0.013$ per cent,
and the ratio $(\lambda_2/\lambda_1)^2$ to $0.02$ per cent. The solver is
therefore accurate in the absolute frequencies as well as in the ratio, and every
beam result below is computed with it.

% ============================================================
\section{Results for beams}
\label{sec:beamresults}

\subsection{Dependence on slenderness}
\label{sec:beamthick}
The shear and rotary-inertia corrections are examined first on their own, by
sweeping the slenderness $h/L$ of the local ($e_0a=0$) Timoshenko beam; this is the
beam counterpart of the plate thickness sweep of Section~\ref{sec:resultsI}.
Table~\ref{tab:beam_thick} reports the spectral ratios of the simply supported,
clamped-clamped and clamped-free beams and the two shifts formed from them. The
absolute ratios all soften by about $13$ per cent between $h/L=0.01$ and $0.20$,
because shear deformation lowers the second mode more than the first. The shift
$\Delta r$(SS$\to$CC), by contrast, stays between $-0.311$ and $-0.318$ across the
whole range, a variation of two per cent of its own value, and
$\Delta r$(CC$\to$CF) between $1.26$ and $1.30$. The asymmetry that
Section~\ref{sec:resultsI} finds for plates therefore holds for beams as well: the
ratio carries the shear correction, the shift largely cancels it.

\begin{table}[t]
\centering
\caption{Spectral ratios of the local Timoshenko beam and the shifts between
boundary conditions as functions of slenderness $h/L$ ($e_0a=0$, $\nu=0.3$,
$\kappa^2=5/6$). The absolute ratios soften by about 13 per cent across the range;
the shifts vary by two to three per cent of their values.}
\label{tab:beam_thick}
\begin{tabular}{cccccc}
\toprule
$h/L$ & $r$(SS) & $r$(CC) & $r$(CF) & $\Delta r$(SS$\to$CC) & $\Delta r$(CC$\to$CF) \\
\midrule
0.01 & 3.9980 & 2.7546 & 6.2640 & $-0.3110$ & $1.2740$ \\
0.02 & 3.9919 & 2.7490 & 6.2554 & $-0.3114$ & $1.2755$ \\
0.05 & 3.9509 & 2.7121 & 6.1964 & $-0.3135$ & $1.2847$ \\
0.10 & 3.8214 & 2.6078 & 6.0025 & $-0.3176$ & $1.3017$ \\
0.15 & 3.6495 & 2.4919 & 5.7264 & $-0.3172$ & $1.2979$ \\
0.20 & 3.4684 & 2.3907 & 5.4119 & $-0.3107$ & $1.2637$ \\
\bottomrule
\end{tabular}
\end{table}

\subsection{Nonlocality, the absolute frequencies and the ratio}
\label{sec:resultsII}

This subsection establishes that the absolute frequencies and the spectral ratio
respond differently to nonlocality. The computations use
the nonlocal Timoshenko model of Section~\ref{sec:nonlocal}, with $e_0a$ swept from
0 to 2~nm.

\begin{figure}[t]
\centering
\includegraphics[width=0.7\textwidth]{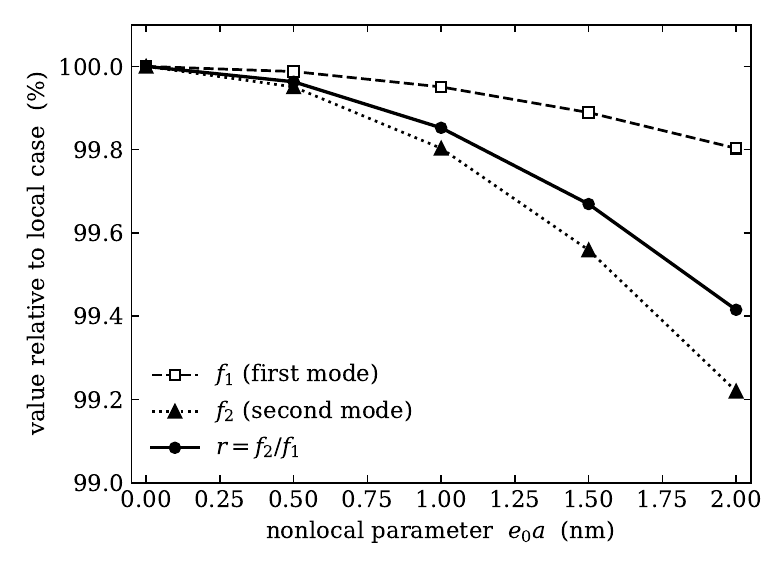}
\caption{Nonlocal softening of the first two natural frequencies and the spectral
ratio, each relative to the local case, for a simply supported SWCNT.}
\label{fig:fig2}
\end{figure}

Table~\ref{tab:nonlocal} reports, for a representative SWCNT of diameter 1.4~nm and
length 100~nm under simply supported conditions, the first two natural frequencies
and their ratio as functions of the nonlocal parameter, with frequencies
normalised to their local values. Both frequencies decrease as the nonlocal
parameter increases, the established nonlocal softening
\citep{Eringen1983,Chwal2018}, and the second mode decreases more than the first:
at $e_0a=2$~nm the fundamental has fallen by 0.20 per cent and the second mode by
0.78 per cent. The ratio changes by 0.59 per cent across the range. Since the
fundamental is almost unchanged, the ratio inherits nearly the whole of the second
mode's shift, so it is somewhat flatter than the second mode, but only modestly so:
it would be an overstatement to call it nonlocal-invariant on the strength of a
single boundary condition. The strong cancellation appears only when the ratio of
one boundary condition is compared with that of another, because the differential
softening is itself nearly the same for different boundary conditions and so
cancels a second time in the shift $\Delta r$. It is the difference of ratios, not
the ratio itself, that suppresses the nonlocal signature, as Section~\ref{sec:robust}
shows. Figure~\ref{fig:fig2} displays the contents of Table~\ref{tab:nonlocal}.

\begin{table}[t]
\centering
\caption{Nonlocal effect on the first two natural frequencies and their ratio for
a simply supported SWCNT ($d=1.4$~nm, $L=100$~nm). Frequencies are relative to the
local ($e_0a=0$) values.}
\label{tab:nonlocal}
\begin{tabular}{ccccc}
\toprule
$e_0a$ (nm) & $f_1$ (rel.) & $f_2$ (rel.) & $r=f_2/f_1$ & shift in $r$ (\%) \\
\midrule
0.0 & 1.0000 & 1.0000 & 3.9940 & $0.000$ \\
0.5 & 0.9999 & 0.9995 & 3.9926 & $-0.037$ \\
1.0 & 0.9995 & 0.9980 & 3.9882 & $-0.148$ \\
1.5 & 0.9989 & 0.9956 & 3.9808 & $-0.331$ \\
2.0 & 0.9980 & 0.9922 & 3.9707 & $-0.585$ \\
\bottomrule
\end{tabular}
\end{table}

\subsection{Robustness of the shift against nonlocality}
\label{sec:robust}
\begin{figure}[t]
\centering
\includegraphics[width=0.7\textwidth]{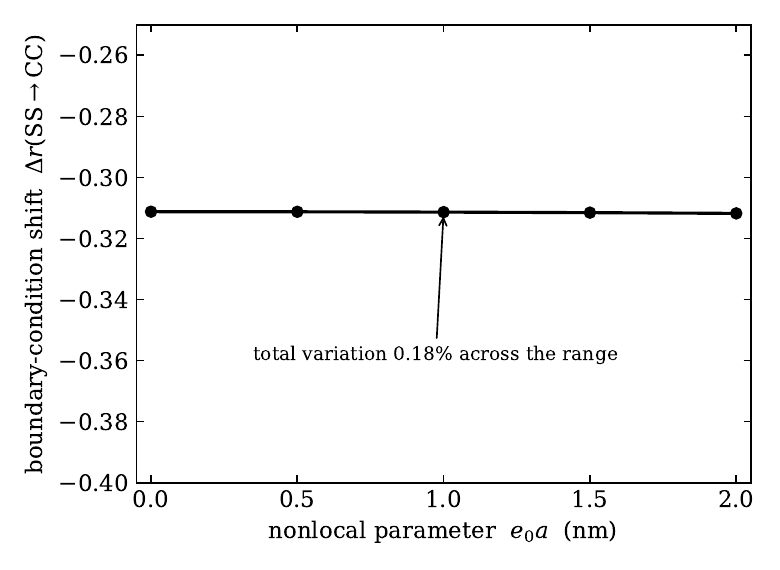}
\caption{Boundary-condition shift $\Delta r$(SS$\to$CC) as a function of the
nonlocal parameter, showing near-invariance across the literature range.}
\label{fig:fig3}
\end{figure}

Table~\ref{tab:robust_nl} reports the spectral ratio of the simply supported and
the doubly clamped nanotube, and the shift between them, as functions of the
nonlocal parameter. Both ratios drift downward by similar fractions, so their
fractional difference is almost unchanged: the shift $\Delta r$(SS$\to$CC) moves
from $-0.3112$ at $e_0a=0$ to $-0.3118$ at $e_0a=2$~nm, a change of 0.2 per cent
across the full literature range. This is the second cancellation anticipated in
Section~\ref{sec:resultsII}: the residual differential softening is itself nearly
common to the two boundary conditions, so it cancels when the two ratios are
differenced. A measurement of $\Delta r$ therefore reports on the boundary
condition without requiring the small-scale parameter to be known.
Figure~\ref{fig:fig3} shows this near-invariance.

\begin{table}[t]
\centering
\caption{Spectral ratios of the simply supported (SS) and doubly clamped (CC)
nanotube and the shift between them, as functions of the nonlocal parameter
($d=1.4$~nm, $L=100$~nm).}
\label{tab:robust_nl}
\begin{tabular}{cccc}
\toprule
$e_0a$ (nm) & $r$(SS) & $r$(CC) & $\Delta r$(SS$\to$CC) \\
\midrule
0.0 & 3.9940 & 2.7510 & $-0.3112$ \\
0.5 & 3.9926 & 2.7498 & $-0.3113$ \\
1.0 & 3.9882 & 2.7464 & $-0.3114$ \\
1.5 & 3.9808 & 2.7406 & $-0.3115$ \\
2.0 & 3.9707 & 2.7327 & $-0.3118$ \\
\bottomrule
\end{tabular}
\end{table}

\subsection{Comparison with published nanoscale data and the regime of validity}
\label{sec:compare}
The robustness results above are computational; this subsection compares the
predicted spectral ratios with independent data and states the regime in which
the comparison is meaningful. The cleanest external check comes from
molecular-dynamics (MD) simulations, in which the boundary conditions are imposed
exactly and no axial tension is present. \citet{Zhang2009} reported MD frequencies
for cantilever single-walled carbon nanotubes and found that the ratio of the
first two flexural modes converges to approximately 6.1 to 6.2 as the aspect ratio
increases, against the Euler--Bernoulli cantilever value of 6.27, the small
shortfall being the expected shear and rotary-inertia correction that the
Timoshenko model supplies. The boundary-condition shift estimated from those data,
$\Delta r \approx (6.15-2.76)/2.76 \approx 1.23$, lies within about three per cent
of the computed value $1.275$ (Section~\ref{sec:inverse}). Because the MD boundary conditions are controlled,
this is a genuine, if approximate, confirmation that the predicted spectral ratio
and its shift hold at the nanoscale. The value of controlled atomistic benchmarks
for verifying nonclassical continuum models of nanobeams has been emphasised
recently by \citet{Darban2025}, who provides molecular-dynamics reference
solutions for the tension, bending, buckling, and free vibration of nanobeams in
clamped and cantilever configurations.

Experimental suspended-nanotube resonators require more care.
\citet{Castellanos2012} measured two flexural modes of a doubly clamped
single-walled nanotube of suspended length 600~nm; the clamped-clamped
Euler--Bernoulli model predicts $f_1=151$~MHz and $f_2=415$~MHz, a ratio of 2.75
that coincides with the present $r(\mathrm{CC})=2.75$, yet the measured frequencies
were $f_A=175$~MHz and $f_B=957$~MHz, a ratio far from 2.76, because the second
detected peak is shifted by residual axial tension and by single-electron effects
specific to that device, and because even modes are weakly transduced in that
geometry. This illustrates the central practical caveat: a real suspended nanotube
is frequently tension-dominated rather than bending-dominated, and in the tension
limit its spectrum approaches that of a vibrating string, $f_n=n f_1$, with
consecutive ratios 2, 3, and so on, rather than the bending values. The spectral
ratios and shifts computed here apply in the bending-dominated regime, in which
the restoring force is set by the flexural rigidity rather than by axial tension
or slack. The diagnostic proposed in this paper is therefore to be applied to
devices, or operating points, for which the bending regime has been established,
for example by confirming the absence of strong gate-voltage tuning of the
resonance frequencies.

That the first-to-second mode ratio is a controllable and meaningful design
quantity is itself established in micro-resonator practice: in
atomic-force-microscopy cantilevers the ratio $f_2/f_1$ is deliberately engineered,
by shaping or by adding mass, across a wide range below the natural
rectangular-cantilever value, precisely because it governs higher-harmonic and
multimode operation \citep{Sadewasser2006}. The present results place this practice
on a common footing with the nanoscale: in every case the ratio is fixed by the
governing operator and the boundary conditions, and the shift $\Delta r$ isolates
that dependence from the scale- and material-dependent corrections.

% ============================================================
\section{Identifying boundary conditions from the shift}
\label{sec:inverse}

The preceding sections established $\Delta r$ as a doubly differenced quantity
that is largely indifferent to scale, constitutive, and material detail. That
property is useful only if it can be put to work, and this section does so by
posing the identification of a boundary condition from measured frequencies as an
inverse problem, characterising the resolution with which $\Delta r$ solves it,
and demonstrating the procedure on independent data. The outcome is a quantitative
method rather than an observation: it states what can be inferred, to what
precision, and under what conditions.

\subsection{The inverse problem and the same-platform principle}
The forward problem maps a boundary condition to a spectrum: given the governing
operator and a boundary-condition set $B$, the eigenproblem yields $r(B)$. The
inverse problem reverses this, inferring $B$ from measured frequencies. Its
difficulty is that the measured ratio depends not on $B$ alone but on a vector of
nuisance parameters $\theta$, comprising the overall scale, the thickness ratio,
the small-scale parameter $e_0a$, and the material constants, several of which are
precisely the quantities that are hard to determine independently in a micro- or
nanoscale device. A single measured ratio therefore confounds the boundary
condition with $\theta$, and recovering $B$ from $r$ requires $\theta$ to be
known.

The boundary-condition shift removes this confound by construction, but only if
the measurement is organised so that the nuisance vector is shared. Suppose the
first two frequencies are measured in two support states of the same structure,
so that $\theta$ is common to both, and form $\Delta r=[r(B_2)-r(B_1)]/r(B_1)$.
Every contribution that enters the two ratios in common cancels in the difference,
and Section~\ref{sec:robust} quantified the residual: less than $0.2$ per cent over
the full literature range of $e_0a$, and of order ten per cent under strong
orthotropy. To that tolerance $\Delta r$ is a function of the boundary-condition
pair $(B_1,B_2)$ alone. We call this the same-platform principle; it is the
experimental counterpart of the double differencing.

It is important to be precise about what the principle does and does not
require, because the point is easily misread. The method does not identify a
boundary condition from a single measurement of an unknown device. What it
identifies is a transition: given two ratio measurements that bracket a
deliberate change to the support, $\Delta r$ tells which of the candidate
transitions in the reference library the change actually realised. Nothing about
the boundary condition need be known in advance. What must be known is that a
change was applied and that it acted on the support alone. The protocol is as
follows. Measure the first two natural frequencies in the initial state and form
$r_1=\omega_2/\omega_1$. Apply the intervention to the support. Measure the first
two frequencies again and form $r_2$. The shift $\Delta r=(r_2-r_1)/r_1$ is then
compared with the library. Four frequencies are measured and two ratios formed;
the identification rests on the single number $\Delta r$.

Three configurations satisfy the same-platform requirement. (i) A known
intervention on one device, for example a cantilever measured free and then with
its tip in contact with a stop, or a beam whose end clamp is tightened between
measurements; the transition is identified, and where the achieved restraint is
partial the continua of Sections~\ref{sec:continuum} and~\ref{sec:tipspring}
convert the measured shift into an estimate of the support compliance. (ii) A
known anchor state: if one of the two states can be established by construction,
such as a free end before a support is applied or a clamp of known stiffness, the
boundary condition of the other state follows from $\Delta r$ directly. (iii) A
same-batch reference device: for structures whose support cannot be altered after
fabrication, such as suspended nanotubes, the comparison may be made against a
companion device from the same fabrication batch with a designed support
condition. The two devices then share thickness, material and small-scale
parameters only to first order, and the residual mismatch enters $\Delta r$ as
additional parameter-induced spread; the invariance results of
Section~\ref{sec:robust} are what make this relaxed form of the principle viable,
since $\Delta r$ is insensitive to precisely the parameters that vary within a
batch.

The intervention must act on the support and on nothing else. A change that also
introduces axial tension, adds mass or alters the effective length shifts the
frequencies through channels that do not cancel in the ratio and must be bounded
separately; the bending-only condition of Section~\ref{sec:compare} applies to
both states. Configuration (i) has an established instance in contact-resonance
atomic force microscopy, in which the same cantilever is measured free and in tip
contact and the contact stiffness is recovered from the shift of its resonances
\citep{Rabe1996,Rabe2000}. The spring there acts translationally at the tip,
which is the case treated in Section~\ref{sec:tipspring}.

\subsection{A reference library and its resolution}
Identification proceeds by comparing a measured $\Delta r$ with a library of
computed reference values, one for each boundary-condition transition of interest.
Two quantities used throughout this section are defined first. The
parameter-induced spread $\delta$ of a transition is the range of its shift over
the nuisance parameter concerned,
\begin{equation}
\delta=\max_{\theta}\Delta r-\min_{\theta}\Delta r,
\end{equation}
evaluated over the full literature range of $e_0a$ for the nanotube and over the
orthotropy contrast of Section~\ref{sec:ortho} for the plate; it is the width of
the band within which a measured shift is assigned to that transition. The
resolution of an identification is the discrimination margin $M$ between the two
nearest transitions, defined below, and two classes are taken as resolved when
$M>1$. Table~\ref{tab:library} assembles the library, by system, from the results
of the preceding sections, with the spread of each entry; Table~\ref{tab:margins}
lists the margins between the transitions of each system.

\begin{table}[t]
\centering
\caption{Reference library of boundary-condition shifts assembled from the present
results, with the parameter-induced spread $\delta$ that sets the identification
band. NT denotes the nanotube (nonlocal Timoshenko) system, with spreads over
$e_0a\in[0,2]$~nm; P the square plate, with spreads over the orthotropy contrast
of Section~\ref{sec:ortho}. The plate reference values are taken at $h/a=0.05$.}
\label{tab:library}
\begin{tabular}{llcl}
\toprule
System & Transition & Reference $\Delta r$ & Dominant nuisance and spread $\delta$ \\
\midrule
NT & SS $\to$ CC       & $-0.311$ & nonlocality, $5.6\times10^{-4}$ ($0.2\%$) \\
NT & CF $\to$ CP       & $-0.483$ & nonlocality, $3.5\times10^{-4}$ ($0.07\%$) \\
NT & CC $\to$ CF       & $+1.275$ & nonlocality, $2.7\times10^{-3}$ ($0.2\%$); atomistic check to $3\%$ \citep{Zhang2009} \\
P  & SSSS $\to$ SCSC   & $-0.221$ & orthotropy, $0.003$ ($1.4\%$) \\
P  & SSSS $\to$ CCCC   & $-0.188$ & orthotropy, $0.019$ ($9.9\%$) \\
\bottomrule
\end{tabular}
\end{table}

\begin{table}[t]
\centering
\caption{Discrimination margins $M$ between the transitions of each system in
Table~\ref{tab:library}, with $\delta$ the larger of the two spreads. All beam
pairs are resolved by two to three orders of magnitude; the finer plate
distinction is resolved with a margin of about 1.4 under strong orthotropy.}
\label{tab:margins}
\begin{tabular}{llcc}
\toprule
System & Pair & Separation $|\Delta r_A-\Delta r_B|$ & Margin $M$ \\
\midrule
NT & SS$\to$CC vs CF$\to$CP   & $0.172$ & $\approx 300$ \\
NT & SS$\to$CC vs CC$\to$CF   & $1.586$ & $\approx 600$ \\
NT & CF$\to$CP vs CC$\to$CF   & $1.758$ & $\approx 650$ \\
P  & SSSS$\to$SCSC vs SSSS$\to$CCCC & $0.033$ & $\approx 1.7$ \\
\bottomrule
\end{tabular}
\end{table}

Whether two boundary-condition classes can be told apart is decided by the ratio
of their separation in $\Delta r$ to the parameter-induced spread, which we call
the discrimination margin,
\begin{equation}
M = \frac{|\Delta r_A - \Delta r_B|}{\delta},
\end{equation}
with $\delta$ the larger of the two spreads. For the beam transitions the margin
is large. The simply supported to clamped transition sits at $\Delta r=-0.311$
with a nonlocal spread of $5.6\times10^{-4}$, the cantilever-to-pinned-tip
transition at $-0.483$ with a spread of $3.5\times10^{-4}$, and the
clamped-to-cantilever transition at $+1.275$ with a spread of $2.7\times10^{-3}$;
the smallest separation among them, $0.172$, divided by the largest spread gives a
margin of about $300$, and the other pairs are resolved more strongly still
(Table~\ref{tab:margins}). Under the larger orthotropic perturbation of the plate
the spread widens to about $0.02$, and the finer distinction between the two
clamped plate transitions, separated by about $0.03$, is then resolved only with a
margin of about $1.7$, whereas the major simply-supported-to-clamped distinction
remains resolved by more than an order of magnitude. The method therefore carries an explicit and honest resolution: it
separates the principal support conditions robustly under every perturbation
examined, and separates finer gradations of constraint when the dominant nuisance
is small-scale rather than material.

The contrast with the naive use of the absolute ratio is what gives the method its
value. Identifying a boundary condition from $r$ itself exposes the inference to
the full nuisance spread of the absolute ratio, which Sections~\ref{sec:resultsI}
and~\ref{sec:robust} measured at $12$ to $14$ per cent across thickness and $6$ to
$9$ per cent across orthotropy. Since the separations between neighbouring
boundary-condition classes in $r$ are themselves of that order, the SCSC and CCCC
plates for instance differing in $r$ by about six per cent, an unknown thickness
or material renders them indistinguishable on the absolute ratio alone. The same
two classes, read through $\Delta r$ on a common platform, are separated by a
quantity from which that nuisance has cancelled. Differencing thus buys roughly an
order of magnitude in identification resolution, and it is this gain, rather than
the robustness of $\Delta r$ considered in the abstract, that constitutes the
practical result.

\subsection{Demonstration}
Two identifications illustrate the procedure, one from independent data and one
synthetic. The molecular-dynamics frequencies of \citet{Zhang2009} for a
cantilevered nanotube give a first-to-second flexural ratio converging to $6.1$ to
$6.2$, so that the shift relative to the doubly clamped reference is
$\Delta r\approx(6.15-2.76)/2.76\approx1.23$, within about three per cent of the
computed value $1.275$ of Table~\ref{tab:library} and, because the atomistic boundary conditions
carry no adjustable small-scale parameter, robust by construction. Read as an
inverse problem, the measured shift assigns the device to the cantilever bending
class without any continuum parameter having been fitted; the identification is
made from the spectrum alone.

For a same-platform example, consider a nanotube whose first two flexural
frequencies are obtained first under simply supported and then under doubly clamped
conditions, giving $r(\mathrm{SS})\approx3.99$ and $r(\mathrm{CC})\approx2.75$ and
hence $\Delta r\approx-0.31$. This value matches the clamped reference of
Table~\ref{tab:library} irrespective of the small-scale parameter, which is unknown
and need not be measured, since the reference shifts by less than one part in a
thousand across its whole range. Had the clamped ratio $r(\mathrm{CC})\approx2.75$
been used on its own, it could not have been separated from, say, a thicker or
orthotropic configuration of higher nominal ratio without independent knowledge of
the thickness and the material; the differenced ratio removes that ambiguity.

\subsection{From classifier to estimator: a partial-clamping continuum}
\label{sec:continuum}
The library of Section~\ref{sec:inverse} treats the boundary condition as one of a
discrete set, but real supports are rarely ideal: a suspended nanotube may be
partially clamped, adhered over a finite length, or otherwise compliant, so that its
constraint lies between the simply supported and clamped extremes. To address this we
extend the nonlocal Timoshenko model of Section~\ref{sec:nonlocal} by replacing the
idealised end conditions with a rotational spring of non-dimensional stiffness
$\kappa_\theta$ at each support, the support's rotational stiffness in units of
$EI/L$, which resists the end rotation through an added energy
$\tfrac12\kappa_\theta[\psi(0)^2+\psi(1)^2]$ while the translational pin is retained.
The same stiffness is assigned to both ends, so the continuum describes equal
partial clamping; unequal end restraint is a straightforward extension with two
stiffness parameters but is not pursued here.
The representation of partial end restraint by rotational springs is a standard
device in the vibration of beams with semirigid supports
\citep{Monsalve2009,Jin2024};
what is new here is the behaviour of $\Delta r$ along the resulting continuum,
rather than the spring model itself.
The limits are recovered: $\kappa_\theta=0$ reproduces the simply supported beam
exactly, and $\kappa_\theta\to\infty$ the clamped beam, the stiff-spring limit
reproducing the clamped spectral ratio of Table~\ref{tab:robust_nl} to within
$0.4$ per cent; the residual is the slow convergence of a sine basis under a stiff
rotational penalty, not a physical difference, and the clamped endpoint of the
table is taken from the direct clamped-clamped computation.

Table~\ref{tab:continuum} reports the shift $\Delta r$ from the simply supported
baseline as a function of $\kappa_\theta$, together with its variation as the
nonlocal parameter is swept from $0$ to $2$~nm. Two features make $\Delta r$ an
estimator of the support stiffness, not merely a classifier of its endpoints. First,
$\Delta r$ is a monotonic function of $\kappa_\theta$, falling smoothly from zero at
the pinned limit to $-0.311$ at the clamped limit, so a measured shift inverts
uniquely to a continuous estimate of the rotational stiffness. Second, at every
stiffness the dependence on the nonlocal parameter stays below $6\times10^{-4}$ in
$\Delta r$, an
order of magnitude smaller than the spacing between successive rows of the table, so
the estimate is obtained without knowledge of the small-scale parameter, exactly as
for the discrete identification. This independence therefore survives the
generalisation: $\Delta r$ measures how clamped a resonator is, not merely whether it
is clamped.

\begin{table}[t]
\centering
\caption{The boundary-condition shift as a continuous estimator of rotational support
stiffness. $\Delta r$ is the shift from the simply supported baseline; the last column
is the change in $\Delta r$ as the nonlocal parameter is swept from $0$ to $2$~nm
($d=1.4$~nm, $L=100$~nm). The pinned and clamped limits agree with the independent
SS and CC computations of Table~\ref{tab:robust_nl}.}
\label{tab:continuum}
\begin{tabular}{cccc}
\toprule
$\kappa_\theta$ & $\Delta r$ ($e_0a=0$) & $\Delta r$ ($e_0a=2$~nm) & nonlocal spread \\
\midrule
$0$ (pinned)       & $0.000$  & $0.000$  & --- \\
$1$                & $-0.108$ & $-0.108$ & $1\times10^{-6}$ \\
$2$                & $-0.166$ & $-0.166$ & $5\times10^{-6}$ \\
$5$                & $-0.245$ & $-0.245$ & $3\times10^{-5}$ \\
$10$               & $-0.287$ & $-0.287$ & $9\times10^{-5}$ \\
$20$               & $-0.309$ & $-0.309$ & $2\times10^{-4}$ \\
$\infty$ (clamped) & $-0.311$ & $-0.312$ & $6\times10^{-4}$ \\
\bottomrule
\end{tabular}
\end{table}

\subsection{A second continuum: the cantilever with a tip support}
\label{sec:tipspring}
The rotational-spring continuum runs between the pinned and clamped conditions of
a doubly supported beam. The intervention most easily realised on a cantilever is
different: bringing the free end into contact with a stop, which adds a
translational restraint at the tip. This is modelled by the translational spring
of Section~\ref{sec:beamdisc}, of stiffness $\kappa_w$ in units of $EI/L^3$ at
$x=1$; $\kappa_w=0$ is the free cantilever, and $\kappa_w\to\infty$ pins the tip
while leaving its rotation free, which is the clamped-pinned beam. The limits
validate the model: the clamped-free and clamped-pinned ratios are recovered to
$0.01$ and $0.004$ per cent (Section~\ref{sec:beamvalid}).

Table~\ref{tab:tipspring} reports $\Delta r$ from the clamped-free baseline along
this continuum. Nonlocal invariance holds throughout, the spread over
$e_0a\in[0,2]$~nm staying below $1.1\times10^{-3}$, and the endpoint transition
CF$\to$CP, $\Delta r=-0.483$ with a spread of $3.5\times10^{-4}$, is the most
invariant entry in the library. But the continuum differs from the rotational one
in an important respect: $\Delta r$ is not monotonic in $\kappa_w$. It falls from
zero to a minimum of $-0.629$ near $\kappa_w\approx63$ and then rises to the
clamped-pinned value $-0.483$. The reason is visible in the individual
frequencies. The tip spring raises the fundamental early and saturates it by
$\kappa_w\approx300$, whereas the second mode, whose own bending stiffness is some
fourteen times larger, barely responds below $\kappa_w\approx20$ and rises
steeply only between $10^{2}$ and $10^{3}$. The two modes respond on different
stiffness scales, so their ratio passes through a minimum; no mode crossing is
involved, the two lowest frequencies never approaching one another. This is the
familiar behaviour of contact-resonance measurement, in which the modes of a
cantilever in tip contact are known to respond to the contact stiffness on
different scales \citep{Rabe1996,Rabe2000}.

The consequence for estimation is stated plainly. A measured shift in the
interval $(-0.629,-0.483]$ corresponds to two values of $\kappa_w$, one on either
side of the minimum, and the ratio alone cannot choose between them; a shift in
$(-0.483,0)$ inverts uniquely to the low-stiffness branch. Resolving the ambiguous
band needs one further piece of information, such as the direction in which the
shift moved as the contact was progressively made, or the absolute fundamental,
which is monotonic in $\kappa_w$ but reintroduces the scale dependence the ratio
was chosen to avoid. The endpoint identification CF$\to$CP is unaffected.

\begin{table}[t]
\centering
\caption{The boundary-condition shift along the tip-support continuum. $\Delta r$
is the shift from the clamped-free baseline; the last column is the change in
$\Delta r$ as the nonlocal parameter is swept from $0$ to $2$~nm ($d=1.4$~nm,
$L=100$~nm). The dependence on $\kappa_w$ is non-monotonic, with a minimum near
$\kappa_w\approx63$.}
\label{tab:tipspring}
\begin{tabular}{cccc}
\toprule
$\kappa_w$ & $\Delta r$ ($e_0a=0$) & $\Delta r$ ($e_0a=2$~nm) & nonlocal spread \\
\midrule
$0$ (free)            & $0.000$  & $0.000$  & --- \\
$1$                   & $-0.126$ & $-0.126$ & $3\times10^{-5}$ \\
$2$                   & $-0.211$ & $-0.211$ & $6\times10^{-5}$ \\
$5$                   & $-0.359$ & $-0.359$ & $1\times10^{-4}$ \\
$10$                  & $-0.473$ & $-0.474$ & $2\times10^{-4}$ \\
$20$                  & $-0.565$ & $-0.566$ & $2\times10^{-4}$ \\
$50$                  & $-0.627$ & $-0.627$ & $8\times10^{-5}$ \\
$75$                  & $-0.628$ & $-0.628$ & $3\times10^{-4}$ \\
$100$                 & $-0.620$ & $-0.619$ & $6\times10^{-4}$ \\
$200$                 & $-0.580$ & $-0.579$ & $1\times10^{-3}$ \\
$500$                 & $-0.526$ & $-0.525$ & $1\times10^{-3}$ \\
$1000$                & $-0.503$ & $-0.503$ & $4\times10^{-4}$ \\
$\infty$ (pinned tip) & $-0.483$ & $-0.484$ & $3.5\times10^{-4}$ \\
\bottomrule
\end{tabular}
\end{table}

\subsection{Procedure and scope}
The method reduces to four steps. First, confirm that the device is in the
bending-dominated regime, for instance by verifying the absence of strong
gate-voltage tuning of the resonance frequencies, since a tension-dominated
resonator follows the string spectrum rather than the bending one
(Section~\ref{sec:compare}). Second, measure the first two flexural frequencies
before and after a deliberate change to the support, in one of the three
configurations of Section~\ref{sec:inverse}. Third, form $\Delta r$. Fourth,
assign the boundary-condition transition by matching $\Delta r$ to the reference
library within its resolution band. The scope is that of the underlying results:
flexural modes of bending-dominated beams and plates, with boundary-condition
pairs drawn from the library. Beyond that library, Sections~\ref{sec:continuum}
and~\ref{sec:tipspring} showed that a measured $\Delta r$ also inverts to a
continuous estimate of the support stiffness, uniquely along the rotational
continuum and, along the tip-support continuum, uniquely outside the band stated
there. What remains for future work is the experimental realisation: a
same-platform measurement on a device verified to lie in the bending regime,
framed explicitly as a measurement of $\Delta r$.

\section{Extension beyond beams: shear-deformable plates}
\label{sec:plates}

The beam results above are one-dimensional. This section extends the same
programme to the two-dimensional case, a shear-deformable square plate, to
establish that the properties of the ratio and the shift are not artefacts of the
beam operator. The plate work is kept self-contained: model, discretisation and
validation are set out first, then the dependence on thickness, then the effect
of material orthotropy, which is the third perturbation considered in this paper
and one that has no one-dimensional counterpart. Nonlocality is not introduced for
the plate; nonlocal plate vibration, including stress-driven formulations, is
treated elsewhere \citep{Jafarinezhad2023,Jafarinezhad2024}, and the plate serves
here to test the
ratio and the shift against the shear and material perturbations, the nonlocal
question having been settled on the beam.

\subsection{First-order shear deformation plate model}
The shear-deformable plate is described by first-order shear deformation theory
\citep{Reissner1945,Mindlin1951}, retaining three independent fields: the
transverse displacement $w(x,y)$ and the cross-section rotations $\psi_x(x,y)$ and
$\psi_y(x,y)$. For a plate of uniform thickness $h$, side length $a$, density
$\rho$, Young's modulus $E$, Poisson ratio $\nu$, and shear correction factor
$\kappa^2$, the bending rigidity is $D=Eh^3/[12(1-\nu^2)]$ and the transverse
shear rigidity is $S=\kappa^2 G h$, with $G=E/[2(1+\nu)]$. Hamilton's principle
gives a coupled eigenvalue problem which, after spatial discretisation, takes the
generalised form
\begin{equation}
\mathbf{K}\,\mathbf{q} = \omega^2 \mathbf{M}\,\mathbf{q},
\end{equation}
with $\mathbf{q}$ the discrete amplitudes of $(w,\psi_x,\psi_y)$, $\mathbf{K}$ the
symmetric stiffness matrix from the bending and shear energies, and $\mathbf{M}$
the consistent mass matrix including translational and rotary inertia. Frequencies
are reported as $\lambda=\omega a^2\sqrt{\rho h/D}$, the convention of
\citet{Leissa1969}, in which the simply supported square plate gives
$\lambda_1=2\pi^2$; note that this differs from the beam convention of
Section~\ref{sec:beam}, where the frequency parameter is defined so that
$\lambda_n=n\pi$ and the reported quantity is $\lambda^2$. The Kirchhoff thin
limit is recovered as $h/a\to 0$, and the shear correction enters through the
non-dimensional shear rigidity $s=\kappa^2\,6(1-\nu)/(h/a)^2$. The shear
correction factor is taken as $\kappa^2=5/6$ throughout.

\subsection{Discretisation and boundary conditions}
The plate eigenvalue problem is solved by the finite element method using the
scikit-fem library \citep{Gustafsson2020}, with the three Mindlin fields
discretised on triangular elements using quadratic (P2) vector basis functions.
Free-vibration analysis of plates under a range of boundary conditions is well
established, through Rayleigh--Ritz, finite element, isogeometric,
integral-transform and dynamic stiffness formulations
\citep{Mei2026,Singh2023,Guo2026}, against which the present
finite element treatment
is a standard and deliberately simple choice; the contribution of this paper lies
in the quantity studied, not in the solver. The square domain is meshed by uniform
refinement, and the smallest eigenvalues of the constrained generalised problem are
extracted by a shift-invert Lanczos iteration. Three boundary-condition sets are
used, all built from simply supported (S) and clamped (C) edges and named by the
condition on the four edges in order: SSSS, SCSC, and CCCC. Free edges are
excluded, because the corner conditions and twisting-moment terms of a free plate
edge require special treatment outside the present scope.

\subsection{Validation}
\label{sec:platevalid}
The plate solver passes two checks. In the thin-plate limit the finite element
Mindlin model is checked against
the classical Kirchhoff results tabulated by \citet{Leissa1969}. The quantity on
which every result of this paper depends is the spectral ratio, and it is the ratio
that the model reproduces accurately. At $h/a=0.005$ the computed ratio
$r=\omega_2/\omega_1$ agrees with the Kirchhoff value $\lambda_2/\lambda_1$ to within
$0.1$ per cent for the simply supported and fully clamped plates and to about $0.7$
per cent for the mixed plate (Table~\ref{tab:valid}). The absolute fundamental
parameter is less accurate: the equal-order element does not fully resolve the shear
constraint at a clamped edge in the thin limit, so the clamped and mixed plates come
out low by of order one per cent, whereas the simply supported plate, which has no
clamped edge, is accurate to about a tenth of a per cent. That the ratio is markedly
more accurate than the absolute frequencies from which it is formed is the first
appearance of the central property of this paper: the spectral ratio cancels the
common-mode part of the discretisation error, just as it later cancels the
common-mode physical contributions.

The equal-order formulation shows shear locking only for the thinnest
plates: at $h/a=0.001$ the fundamental parameters rise two to four per cent above the
benchmark, but this overstiffening has disappeared by $h/a=0.005$, the thickness at
which the check above is performed. The study proper is conducted in the range
$0.01\le h/a\le 0.20$, which is where first-order shear deformation is the
physically appropriate model; thinner plates are accurately described by classical
thin-plate theory.

\begin{table}[t]
\centering
\caption{Validation of the finite element Mindlin plate against the Kirchhoff
benchmarks of \citet{Leissa1969} in the thin limit ($h/a=0.005$, square isotropic
plate, $\nu=0.3$). The spectral ratio $r=\omega_2/\omega_1$ is reproduced far more
accurately than the absolute fundamental $\lambda_1$, the common-mode element error
cancelling in the ratio.}
\label{tab:valid}
\begin{tabular}{lccccc}
\toprule
BC & $\lambda_1$ (present) & $\lambda_1$ (Leissa) & $r$ (present) & $r$ (Leissa) & $r$ error (\%) \\
\midrule
SSSS & 19.757 & 19.739 & 2.5008 & 2.5000 & $+0.03$ \\
SCSC & 28.752 & 28.951 & 1.9043 & 1.8909 & $+0.71$ \\
CCCC & 35.656 & 35.992 & 2.0413 & 2.0397 & $+0.08$ \\
\bottomrule
\end{tabular}
\end{table}

\subsection{The spectral-shift family and its dependence on thickness}
\label{sec:resultsI}

This section reports the spectral ratio and its boundary-condition shift for the
shear-deformable square plate across the physically relevant range of thickness
ratios, establishing that the spectral ratio is a genuine function of thickness
through the shear correction, not a fixed number.

\begin{figure}[t]
\centering
\includegraphics[width=0.7\textwidth]{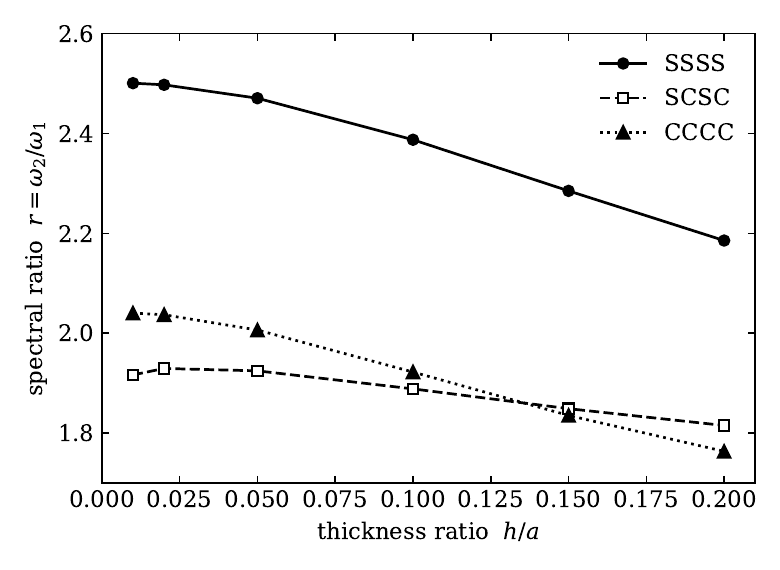}
\caption{Spectral ratio $r=\omega_2/\omega_1$ of the square isotropic plate as a
function of thickness ratio $h/a$ for three boundary conditions.}
\label{fig:fig1}
\end{figure}

Table~\ref{tab:r_thick} reports the spectral ratio $r=\omega_2/\omega_1$ for the
three boundary conditions across $0.01\le h/a\le 0.20$. The ratio decreases with
increasing thickness for the SSSS and CCCC plates, by 12.6 and 13.6 per cent
respectively, as transverse shear softens the higher mode more than the
fundamental, while the SCSC plate is almost insensitive to thickness, its ratio
remaining close to 1.9 throughout. The magnitude, and not only the presence, of
the shear effect on the ratio is therefore itself boundary-condition dependent.
Figure~\ref{fig:fig1} displays these curves.

\begin{table}[t]
\centering
\caption{Spectral ratio $r=\omega_2/\omega_1$ of the square isotropic plate as a
function of thickness ratio $h/a$ ($\nu=0.3$, $\kappa^2=5/6$).}
\label{tab:r_thick}
\begin{tabular}{cccc}
\toprule
$h/a$ & SSSS & SCSC & CCCC \\
\midrule
0.01 & 2.5005 & 1.9167 & 2.0404 \\
0.02 & 2.4972 & 1.9296 & 2.0370 \\
0.05 & 2.4704 & 1.9248 & 2.0065 \\
0.10 & 2.3873 & 1.8887 & 1.9224 \\
0.15 & 2.2850 & 1.8491 & 1.8356 \\
0.20 & 2.1856 & 1.8155 & 1.7637 \\
\bottomrule
\end{tabular}
\end{table}

The shift $\Delta r$ relative to the SSSS baseline is reported in
Table~\ref{tab:dr_thick}. Both shifts are negative, since the more constrained
plates have a lower spectral ratio than the simply supported plate. The important
observation is that the shift is in every case a more slowly varying quantity than
the absolute ratios from which it is formed, foreshadowing the central result of
Section~\ref{sec:robust}.

\begin{table}[t]
\centering
\caption{Boundary-condition spectral shift $\Delta r$ relative to the SSSS
baseline, as a function of thickness ratio $h/a$.}
\label{tab:dr_thick}
\begin{tabular}{ccc}
\toprule
$h/a$ & $\Delta r$(SSSS$\to$SCSC) & $\Delta r$(SSSS$\to$CCCC) \\
\midrule
0.01 & $-0.2335$ & $-0.1840$ \\
0.02 & $-0.2273$ & $-0.1843$ \\
0.05 & $-0.2208$ & $-0.1878$ \\
0.10 & $-0.2088$ & $-0.1947$ \\
0.15 & $-0.1907$ & $-0.1967$ \\
0.20 & $-0.1693$ & $-0.1930$ \\
\bottomrule
\end{tabular}
\end{table}

% ============================================================

\subsection{Robustness of the shift against material orthotropy}
\label{sec:ortho}
The square plate of Section~\ref{sec:resultsI} is recomputed with orthotropic
bending properties representative of a layered material, with
$D_{22}/D_{11}=0.6$, and compared with the isotropic case at fixed thickness
(Table~\ref{tab:ortho}). The absolute spectral ratio changes by 6 to 9 per cent
across the three boundary conditions, since the anisotropy alters the relative
stiffness of the modes. The boundary-condition shift is far less affected:
$\Delta r$(SSSS$\to$SCSC) changes from $-0.217$ to $-0.220$ and
$\Delta r$(SSSS$\to$CCCC) from $-0.191$ to $-0.172$, in both cases stable to within
about ten per cent, an order of magnitude less sensitive than the absolute ratio.

\begin{table}[t]
\centering
\caption{Effect of material orthotropy on the spectral ratio and the
boundary-condition shift, square plate at $h/a=0.05$, orthotropic case
$D_{22}/D_{11}=0.6$.}
\label{tab:ortho}
\begin{tabular}{lccc}
\toprule
Quantity & Isotropic & Orthotropic & change (\%) \\
\midrule
$r$(SSSS) & 2.446 & 2.235 & $-8.6$ \\
$r$(SCSC) & 1.916 & 1.744 & $-9.0$ \\
$r$(CCCC) & 1.979 & 1.850 & $-6.5$ \\
$\Delta r$(SSSS$\to$SCSC) & $-0.217$ & $-0.220$ & $+1.4$ \\
$\Delta r$(SSSS$\to$CCCC) & $-0.191$ & $-0.172$ & $-9.9$ \\
\bottomrule
\end{tabular}
\end{table}

\subsection{Synthesis across beams and plates}
Across all three perturbations, and for beams and plates alike, the absolute
spectral ratio responds substantially: by about 13 per cent to beam slenderness
and 12 to 14 per cent to plate thickness, by 6 to 9 per cent to orthotropy, and
measurably to nonlocality in the single-boundary-condition sense of
Section~\ref{sec:resultsII}. The boundary-condition shift $\Delta r$, formed from
the same ratios, responds far less in every case: by two per cent of its value
across the beam slenderness range, and most strikingly against nonlocality, where
it changes by less than 0.2 per cent. Each perturbation enters the
absolute frequencies in a largely common-mode way, so it cancels in the ratio to
leading order and cancels again in the difference of ratios between boundary
conditions. The shift is therefore a doubly differenced quantity, and the
corrections that contaminate the absolute frequencies are suppressed at each
differencing, leaving a fingerprint of the boundary condition. The same mechanism
appeared, independently of any physical effect, in the plate validation of
Section~\ref{sec:platevalid}, where the ratio cancelled a purely numerical
common-mode element error, which indicates that the cancellation is a structural property of the ratio
rather than a coincidence of the particular physics.

% ============================================================
\section{Discussion}
\label{sec:discuss}

\subsection{Why the shift is scale-invariant}
Each physical correction considered here, shear deformation, nonlocality, and
orthotropy, enters the eigenfrequencies in a way dominated by a common-mode
factor, with only a small differential part distinguishing one mode from another.
The ratio cancels the common-mode part, and the difference of ratios between two
boundary conditions cancels most of what the differential part contributes in
common to the two conditions. The boundary-condition shift is, in this sense, a
doubly differenced quantity, and its robustness is the cumulative effect of the
two cancellations. This explains why the shift is more stable against nonlocality,
where the variation is below 0.2 per cent, than against thickness or orthotropy,
where it reaches a few to about ten per cent: nonlocality in the relevant range is a small
perturbation whose differential part is nearly identical for the two boundary
conditions, so the second cancellation is almost complete, whereas thickness and
orthotropy are larger perturbations whose differential parts differ more between
boundary conditions.

\subsection{Limitations}
Five limitations should be stated. First, the plate analysis is restricted to
boundary conditions built from simply supported and clamped edges; free edges,
which introduce corner conditions and twisting-moment terms, are left to future
work, and nonlocality is not introduced for the plate. Second, the nanotube is
represented by a one-dimensional nonlocal Timoshenko model with the cross-section
entering only through the radius of gyration, which is appropriate for the
flexural modes that define the spectral ratio but does not capture
circumferential or shell-like modes, and assumes a single-walled, defect-free,
straight tube. Third, as set out in Section~\ref{sec:compare}, the prediction
applies in the bending-dominated regime; the controlled molecular-dynamics data
confirm it to within about three per cent, whereas experimental suspended-nanotube
devices are often tension-dominated and must be operated, or selected, in the
bending regime for the diagnostic to apply. Fourth, the tip-support continuum of
Section~\ref{sec:tipspring} does not invert uniquely over its whole range: shifts
in the band $(-0.629,-0.483]$ correspond to two tip stiffnesses, and resolving
them needs information beyond the ratio. The rotational continuum has no such
band, and the endpoint identifications are unaffected.

Fifth, two effects present in real resonators but absent from the conservative
eigenvalue model bear on the measurement. A lightly damped resonator of quality
factor $Q$ has damped natural frequencies $\omega_d=\omega_n\sqrt{1-1/(4Q^2)}$, so
the spectral ratio is displaced only at order $Q^{-2}$, and not at all when the
two modes share a quality factor; for the high-$Q$ resonators to which the method
applies, $Q\gtrsim10^2$, this displacement is below $10^{-4}$, well under the
shifts the method resolves, so damping, whether viscous or structural, is a
further common-mode effect that the ratio suppresses. Geometric nonlinearity
makes the resonance amplitude-dependent through mid-plane stretching, so the
measurement must be made at small amplitude, in the linear regime; this is the
same requirement, in another form, as remaining in the bending rather than the
tension-dominated regime. The prediction has not been tested against a dedicated
experiment. A same-platform measurement in one of the three configurations of
Section~\ref{sec:inverse}, verified to lie in the bending regime and framed
explicitly as a measurement of $\Delta r$, would provide that test and is the
natural experimental continuation of this work.

% ============================================================
\section{Conclusion}
\label{sec:concl}

This paper has studied the spectral ratio $r=\omega_2/\omega_1$ and the fractional
shift $\Delta r$ it undergoes when a boundary condition is changed, for nonlocal
Timoshenko beams and for shear-deformable plates. Using a locking-free
Rayleigh--Ritz beam solver validated against an exact closed-form solution, the
Euler--Bernoulli thin-beam limits and the isogeometric benchmark of
\citet{Lee2013}, and a finite element Mindlin plate solver whose spectral ratios
reproduce the Leissa thin-plate benchmarks to better than 0.1 per cent, four
results were obtained. First, for beams the absolute ratio softens by about 13 per
cent with slenderness and carries the full nonlocal signature, while the shift
varies by two per cent across the slenderness range and by less than 0.2 per cent
across the full literature range of the nonlocal parameter; the ratio and the
shift carry distinct physical information. Second, this robustness was developed
into an inverse method: which boundary-condition transition a deliberate change of
support has produced is identified by differencing two same-platform ratio
measurements into $\Delta r$ and matching it to a computed reference library, with
a resolution that is characterised explicitly and that exceeds, by roughly an
order of magnitude, what the absolute ratio permits when scale and material are
unknown; the method assigns the boundary-condition class of a molecular-dynamics
benchmark from its spectrum alone. Third, the library was extended to two continua
of real supports. Along the rotational-spring continuum the shift inverts uniquely
to the support stiffness; along the tip-support continuum it passes through a
minimum, so that shifts in a stated band correspond to two stiffnesses, while the
nonlocal invariance holds along both. Fourth, the same asymmetry holds for
shear-deformable plates: the absolute ratio responds to thickness by 12 to 14 per
cent and to material orthotropy by 6 to 9 per cent, while the shift stays within
about ten per cent under each. The unifying picture is that $\Delta r$ is formed by
two successive differences that each cancel a common-mode part of the frequency
response, leaving a quantity that tracks the boundary conditions while staying
largely independent of scale, material, and constitutive detail. Its experimental
realisation through a controlled same-platform measurement is the natural next
step.

% ============================================================
\section*{Declaration of competing interest}
The author declares no competing interests.

\section*{Data availability}
The code that reproduces all numerical results and figures will be made available
as supplementary material upon publication.

\section*{Declaration of AI-assisted technologies in the manuscript preparation process}
During the preparation of this study, the author utilized Claude (Anthropic) in order
to improve the language and clarity of the text, assist in writing and debugging the 
code used to reproduce numerical results. 

% ============================================================
\bibliographystyle{elsarticle-num}

\begin{thebibliography}{00}

\bibitem[Madinier et al.(2025)]{Madinier2025}
Madinier, N., Lecl\`{e}re, Q., Ege, K., Berry, A., 2025. Complex dynamic stiffness
identification of panels using inverse methods based on optical deflectometry
measurements. Journal of Sound and Vibration, article 119373.

\bibitem[Ewins(2000)]{Ewins2000}
Ewins, D.J., 2000. Modal Testing: Theory, Practice and Application, 2nd Edition.
Research Studies Press.

\bibitem[Blevins(1979)]{Blevins1979}
Blevins, R.D., 1979. Formulas for Natural Frequency and Mode Shape. Van Nostrand
Reinhold, New York.

\bibitem[Reissner(1945)]{Reissner1945}
Reissner, E., 1945. The effect of transverse shear deformation on the bending of
elastic plates. Journal of Applied Mechanics 12 (2), A69--A77.

\bibitem[Mindlin(1951)]{Mindlin1951}
Mindlin, R.D., 1951. Influence of rotatory inertia and shear on flexural motions
of isotropic elastic plates. Journal of Applied Mechanics 18 (1), 31--38.

\bibitem[Eringen(1972)]{Eringen1972}
Eringen, A.C., 1972. Linear theory of nonlocal elasticity and dispersion of plane
waves. International Journal of Engineering Science 10 (5), 425--435.

\bibitem[Eringen(1983)]{Eringen1983}
Eringen, A.C., 1983. On differential equations of nonlocal elasticity and
solutions of screw dislocation and surface waves. Journal of Applied Physics 54
(9), 4703--4710.

\bibitem[Peddieson et al.(2003)]{Peddieson2003}
Peddieson, J., Buchanan, G.R., McNitt, R.P., 2003. Application of nonlocal
continuum models to nanotechnology. International Journal of Engineering Science
41 (3--5), 305--312.

\bibitem[Tuna and Kirca(2016)]{Tuna2016}
Tuna, M., Kirca, M., 2016. Exact solution of Eringen's nonlocal integral model for
bending of Euler--Bernoulli and Timoshenko beams. International Journal of
Engineering Science 105, 80--92.

\bibitem[Han et al.(2025)]{Han2025}
Han, K., Yu, J., Zhou, H., 2025. Applicability of scale parameters in nonlocal
shear deformation circular copper plates for axisymmetric vibrations. Journal of
Sound and Vibration 612, 119173.

\bibitem[Chwa\l(2018)]{Chwal2018}
Chwa\l, M., 2018. Nonlocal analysis of natural vibrations of carbon nanotubes.
Journal of Materials Engineering and Performance 27, 6087--6096.

\bibitem[Bocko and Lengvarsk\'y(2014)]{Bocko2014}
Bocko, J., Lengvarsk\'y, P., 2014. Bending vibrations of carbon nanotubes by using
nonlocal theory. Procedia Engineering 96, 21--27.

\bibitem[Murmu and Pradhan(2010)]{Murmu2023}
Murmu, T., Pradhan, S.C., 2010. Vibration and buckling analysis of nano-scale
beams via nonlocal elasticity and Timoshenko beam theory: a differential
quadrature approach. Journal of Aerospace Sciences and Technologies, 40--54.

\bibitem[Xu et al.(2025)]{Xu2025}
Xu, X.J., Bu, M.J., Wang, C.H., 2025. Relationships of a micro-structured beam
system and a two-phase nonlocal beam. Archive of Applied Mechanics 95 (10), 228.

\bibitem[Dang et al.(2026)]{Dang2026}
Dang, N.D., Nguyen, D.K., Bui, T.T.H., Le, C.I., 2026. Hierarchical beam element
for nonlinear vibration of porous-core FGM sandwich beams with influence of
homogenization models. Archive of Applied Mechanics 96 (7), 148.

\bibitem[Zhang et al.(2009)]{Zhang2009}
Zhang, Y.Y., Wang, C.M., Tan, V.B.C., 2009. Assessment of Timoshenko beam models
for vibrational behaviour of single-walled carbon nanotubes using molecular
dynamics. Advances in Applied Mathematics and Mechanics 1 (1), 89--106.

\bibitem[Sazonova et al.(2004)]{Sazonova2004}
Sazonova, V., Yaish, Y., \"{U}st\"{u}nel, H., Roundy, D., Arias, T.A., McEuen,
P.L., 2004. A tunable carbon nanotube electromechanical oscillator. Nature 431,
284--287.

\bibitem[Moser et al.(2014)]{Moser2014}
Moser, J., G\"{u}ttinger, J., Eichler, A., Esplandiu, M.J., Liu, D.E., Dykman,
M.I., Bachtold, A., 2014. Nanotube mechanical resonators with quality factors of
up to 5 million. Nature Nanotechnology 9, 1007--1011.

\bibitem[Lee and Park(2013)]{Lee2013}
Lee, S.J., Park, K.S., 2013. Vibrations of Timoshenko beams with isogeometric
approach. Applied Mathematical Modelling 37 (22), 9174--9190.

\bibitem[Leissa(1969)]{Leissa1969}
Leissa, A.W., 1969. Vibration of Plates, NASA SP-160. National Aeronautics and
Space Administration.

\bibitem[Challamel et al.(2024)]{Challamel2024}
Challamel, N., El-Borgi, S., Trabelssi, M., Reddy, J.N., 2024. Free vibration
response of micromorphic Timoshenko beams. Journal of Sound and Vibration 591,
118602.

\bibitem[Timtaoucine et al.(2026)]{Timtaoucine2026}
Timtaoucine, M.E.H., Fantuzzi, N., Qaderi, S., Derradji, M., Fabbrocino, F., 2026.
Vibrations of lattice nanobeams in strain gradient elasticity. European Journal of
Mechanics - A/Solids, article 106054.

\bibitem[Jin et al.(2024)]{Jin2024}
Jin, Y., Lu, Y., Yang, D., Zhao, F., Luo, X., Zhang, P., 2024. An analytical
method for vibration analysis of multi-span Timoshenko beams under arbitrary
boundary conditions. Archive of Applied Mechanics 94 (3), 529--553.

\bibitem[Darban(2025)]{Darban2025}
Darban, H., 2025. MD benchmarks: size-dependent tension, bending, buckling, and
vibration of nanobeams. International Journal of Mechanical Sciences 296, 110316.

\bibitem[Castellanos-Gomez et al.(2012)]{Castellanos2012}
Castellanos-Gomez, A., Meerwaldt, H.B., Venstra, W.J., van der Zant, H.S.J.,
Steele, G.A., 2012. Strong and tunable mode coupling in carbon nanotube
resonators. Physical Review B 86 (4), 041402(R).

\bibitem[Sadewasser et al.(2006)]{Sadewasser2006}
Sadewasser, S., Villanueva, G., Plaza, J.A., 2006. Modified atomic force
microscopy cantilever design to facilitate access of higher modes of oscillation.
Review of Scientific Instruments 77 (7), 073703.

\bibitem[Rabe et al.(1996)]{Rabe1996}
Rabe, U., Janser, K., Arnold, W., 1996. Vibrations of free and surface-coupled
atomic force microscope cantilevers: theory and experiment. Review of Scientific
Instruments 67 (9), 3281--3293.

\bibitem[Rabe et al.(2000)]{Rabe2000}
Rabe, U., Amelio, S., Kester, E., Scherer, V., Hirsekorn, S., Arnold, W., 2000.
Quantitative determination of contact stiffness using atomic force acoustic
microscopy. Ultrasonics 38 (1--8), 430--437.

\bibitem[Monsalve-Cano and Aristiz\'{a}bal-Ochoa(2009)]{Monsalve2009}
Monsalve-Cano, J.F., Aristiz\'{a}bal-Ochoa, J.D., 2009. Stability and free vibration
analyses of an orthotropic singly symmetric Timoshenko beam-column with
generalized end conditions. Journal of Sound and Vibration 328 (4--5), 467--487.

\bibitem[Jafarinezhad et al.(2023)]{Jafarinezhad2023}
Jafarinezhad, M., Sburlati, R., Cianci, R., 2023. Static and free vibration
analysis of functionally graded annular plates using stress-driven nonlocal
theory. European Journal of Mechanics - A/Solids 99, 104955.

\bibitem[Jafarinezhad et al.(2024)]{Jafarinezhad2024}
Jafarinezhad, M., Sburlati, R., Cianci, R., 2024. Nonlocal stress-driven model for
functionally graded Mindlin annular plate: bending and vibration analysis.
Archive of Applied Mechanics 94 (5), 1313--1333.

\bibitem[Gustafsson and McBain(2020)]{Gustafsson2020}
Gustafsson, T., McBain, G.D., 2020. scikit-fem: A Python package for finite
element assembly. Journal of Open Source Software 5 (52), 2369.

\bibitem[Mei et al.(2026)]{Mei2026}
Mei, S., Caprani, C., Cantero, D., 2026. A separation-of-variable dynamic
stiffness method for the free vibration of thin orthotropic rectangular plates
with general homogeneous boundary conditions. Journal of Sound and Vibration,
article 119639.

\bibitem[Singh et al.(2023)]{Singh2023}
Singh, D., Kiran, R., Vaish, R., 2023. Vibration and buckling analysis of
agglomerated CNT composite plates via isogeometric analysis using non-polynomial
shear deformation theory. European Journal of Mechanics - A/Solids 98, 104892.

\bibitem[Guo and Zhang(2026)]{Guo2026}
Guo, H., Zhang, K., 2026. Vibration and buckling analysis of ribbed orthotropic
cantilever Mindlin plates using finite integral transform method. Archive of
Applied Mechanics 96 (1), 31.

\end{thebibliography}

\end{document}